\documentclass[twocolumn]{aastex6}

\newcommand{\htwo}{H$_2$}

\newcommand{\ms}{M$_{\odot}$}
\newcommand{\kms}{$\,\rm km\,s^{-1}$}

\newcommand{\eqq}{\!=\!}

\newcommand{\jone}{{$J\eqq$ 1$-$0}} 
\newcommand{\jtwo}{{$J\eqq$ 2$-$1}}
\newcommand{\jthree}{{$J\eqq$ 3$-$2}}
\newcommand{\jfour}{{$J\eqq$ 4$-$3}}
\newcommand{\jfive}{{$J\eqq$ 5$-$4}}
\newcommand{\jsix}{{$J\eqq$ 6$-$5}}
\newcommand{\jseven}{{$J\eqq$ 7$-$6}}
\newcommand{\jeight}{{$J\eqq$ 8$-$7}}

\newcommand{\jten}{{$J\eqq$ 10$-$9}}

\newcommand{\jthirteen}{{$J\eqq$ 13$-$12}}

\newcommand{\tkin}{$T_{\rm kin}$}
\newcommand{\nhtwo}{$n_{\rm H_2}$}
\newcommand{\Nco}{$N_{\rm CO}$}

\slugcomment{}

\shorttitle{Recovering Physical Properties from CO SLED Modeling}
\shortauthors{Kamenetzky et al.}

\usepackage{color}
\usepackage{natbib}
\usepackage{bm}
\usepackage{hyperref}
\usepackage{listings}
\usepackage{mathrsfs}

\begin{document}

\title{Recovering the Physical Properties of Molecular Gas in Galaxies from CO SLED Modeling} 
\author{J. Kamenetzky\altaffilmark{1}, G. C. Privon\altaffilmark{2,3,4},
  D. Narayanan\altaffilmark{4,5}} \altaffiltext{1}{Westminster College,
  1840 S 1300 E, Salt Lake City, UT 84105 USA}
\altaffiltext{2}{Instituto de Astrof\'isica, Facultad de F\'isica,
  Pontificia Universidad Cat\'olica de Chile, Casilla 306, Santiago
  22, Chile} \altaffiltext{3}{Departamento de Astronom\'ia,
  Universidad de Concepci\'on, Casilla 160-C, Concepci\'on, Chile}
\altaffiltext{4}{Department of Astronomy, University of Florida, 211
  Bryant Space Sciences Center, Gainesville, FL 32611}
\altaffiltext{5}{Cosmic Dawn Center (DAWN), Niels Bohr Institute, University of Copenhagen, Juliane Maries vej 30, DK-2100 Copenhagen, Denmark}
\email{jkamenetzky@westminstercollege.edu}

\begin{abstract}
Modeling of the spectral line energy distribution (SLED) of the CO molecule can reveal the physical conditions (temperature, density) of molecular gas in Galactic clouds and other galaxies. Recently, the {\it Herschel} Space Observatory and ALMA have offered, for the first time, a comprehensive view of the rotational \jfour\ through \jthirteen\  lines, which arise from a complex, diverse range of physical conditions that must be simplified to one, two, or three components when modeled. Here we investigate the recoverability of physical conditions from SLEDs produced by galaxy evolution simulations containing a large dynamical range in physical properties. These simulated SLEDs were generally fit well by one component of gas whose properties largely resemble or slightly underestimate the luminosity-weighted properties of the simulations when clumping due to non-thermal velocity dispersion is taken into account. If only modeling the first three rotational lines, the median values of the marginalized parameter distributions better represent the luminosity-weighted properties of the simulations, but the uncertainties in the fitted parameters are nearly an order of magnitude, compared to approximately $0.2$ dex in the ``best-case" scenario of a fully sampled SLED through \jten. This study demonstrates that while common CO SLED modeling techniques cannot reveal the underlying complexities of the molecular gas, they can distinguish bulk luminosity-weighted properties that vary with star formation surface densities and galaxy evolution, if a sufficient number of lines are detected and modeled.
\end{abstract}

\keywords{galaxies: ISM -- ISM: molecules -- submillimeter}

\section{Introduction}\label{sec:intro}

$^{12}$CO (hereafter, CO) serves as a tracer of cool molecular gas
because of its high-dipole moment and low-lying rotational energy
levels \citep[e.g.,][and references therein]{Bolatto2013,Carilli2013}.  The ratios of its line emission allows us to determine the
physical conditions (temperature, density) of the molecular gas \citep[e.g.,][]{Weiss2007,Sliwa2012,Bayet2013,Meijerink2013,Schirm2014,Greve2014,Spilker2014,Papadopoulos2014,Xu2015,Rosenberg2015,Daddi2015,Kamenetzky2017,Strandet2017} and
the absolute values of the emission (namely of \jone) allows us to
determine the total molecular gas mass \citep[e.g.,][and references therein]{Bolatto2013}.  Combined, this knowledge
allows us to comment on the processes exciting the gas and therefore
its relationship to star formation and galaxy evolution.

The {\it Herschel} Space Observatory opened a new observational window
from 60 to 670 microns that allowed the study of spectral line energy
distributions (SLEDs) through very high-J lines, thanks to the SPIRE
Fourier Transform Spectrometer (FTS) and PACS.  Prior to this, only
the first few rotational transitions of CO were available to be
studied through the atmosphere. Low-J CO, especially \jone, is a well
used tracer of total cold molecular gas, due to its low energy spacing
(the $J_{\rm upper}=1$ level is 5.53 K above ground) and strong dipole
moment.  The CO \jone\ line is generally optically thick and in local
thermodynamic equilibrium (LTE).  As one climbs up the CO ladder to
measure emission from higher-$J$ lines in a cloud or galaxy's SLED, 
the lines begin to fall from LTE
(i.e. when $h\nu>>kT$ as $E(J+1)-E(J)$ becomes larger with higher $J$), 
and non-LTE calculations of
the level populations, optical depth, and resulting emission are
required using a large velocity gradient (LVG) code like RADEX
\citep{VanderTak2007}.

Because the temperature and density are degenerate parameters, a full
examination of the parameter space using a grid method, Monte Carlo
Markov Chain (MCMC), or nested sampling algorithm is important to
characterize the shape of the parameter space and the uncertainty in
any given parameter.  The line luminosities of the \jfour\ through
\jthirteen\ lines of CO, available for local galaxies with SPIRE, were
discovered to be much more luminous than would be predicted by
extrapolating the cold gas emission to higher-J lines, leading
observers to often invoke a second, warmer component of gas to explain
the luminous emission (see \citet{Kamenetzky2017} and references
therein).  The physical condition of the gas (as determined by the
relative luminosities of the lines) is instructive to study as both
the raw material for star formation and as indicating the effects of
star formation via feedback such as radiative and turbulent
excitation.

\citet{Tunnard2016} studied the recoverability of physical conditions using the Large Velocity Gradient (LVG) code 
RADEX and two methods of $\chi^2$ minimization: grid and Monte Carlo Markov Chain (MCMC). 
In essence, they asked, ``If one produces a SLED in RADEX using a 
given set of physical conditions, can one fit the SLED and recover those original physical parameters?" 
They found that the parameters (temperature and density of the colliding 
partner, \htwo) are only recovered to within half a dex, given the degeneracy 
between the parameters and some uncertainty in the modeled SLED (as one would have observationally). 
Including isotopologue lines with isotopologue abundance ratio as a free parameter improves the constraints. 
\citet{Leroy2017} studied the ability of dense gas tracers (HCN, HCO$^+$, HNC, CS) and the first few lines of CO to 
distinguish changes in the dense gas fraction and median volume density for modeled emission from an ensemble of 
gas clouds with log-normal and power law density distributions. 
Their ensembles were combined one-zone models of molecular emission, with a specified distribution 
of densities, but all gas was taken to be isothermal and with a fixed optical depth (and therefore 
escape probability.
Analytical models, however, are often limited in 
their ability to approximate the diverse range of conditions in integrated extragalactic observations, typically assuming, e.g. isothermal or isobaric conditions.

Galaxies are complex, with a diverse range of physical
  conditions, and in principle the observed SLED is a superposition of
  the sum of individual SLEDs originating from all of the CO-emitting
  gas in a galaxy.  What has been missing, therefore, is an
  investigation into the recoverability of physical conditions from
  SLEDs using {\it bona fide} galaxy evolution simulations that
  contain a large dynamic range in physical properties.  Here, we seek
  to do just that.   \citet{Narayanan2014} used smoothed particle
hydrodynamics to perform idealized simulations of isolated and
interacting galaxies, and then produced galaxy-integrated CO SLEDs
given the physical conditions of the simulated gas.  In
  this paper, we investigate whether fitting these theoretical SLEDs
  in the same manner as is done for typical observations recovers the average physical
  conditions in the gas.    In Section \ref{sec:methods} we describe
the methods used to the produce the simulated galaxies, the simulated
SLEDs, and the fitting of those SLEDs.  Our results and conclusions
are described in Sections \ref{sec:results}, \ref{sec:discussion}, and \ref{sec:conclusions}.

\section{Methods}\label{sec:methods}

\subsection{Galaxy Evolution Simulations}

Our basic strategy is to fit model SLEDs from theoretical simulations,
and compare the derived physical properties from these SLEDs to the
actual gas physical properties from the simulations.  The model SLEDs
were derived in \citet{Narayanan2014}, and we defer the reader to that
paper alongside \citet{Narayanan2017} for details regarding the hydrodynamic and radiative transfer
simulations, though summarize the salient points here.

Following \citet{narayanan11b,narayanan12a}, we employ {\sc
    gadget-3} \citep{springel02a,springel03a,springel05a} hydrodynamic
  simulations of idealized galaxies in evolution.
  The galaxies are initialized as exponential disks following the
  \citet{mo98a} formalism, and reside in live \citet{hernquist90a}
  dark matter halos.  The gas is initialized as primordial, and metals
  form as the simulations evolve.  The interstellar medium is modeled
  as multiphase, with clouds pressure-confined by hot ISM
  \citep{mckee77a}.  Star formation proceeds in this cold gas
  following a volumetric \citet{schmidt59a} star formation relation
  with index $N=1.5$ \citep{kennicutt98a,kennicutt12a}.  The ISM is
  pressurized via supernovae via an effective equation of state; here,
  we assume a modest pressurization $q_{\rm EOS} = 0.25$
  \citep{springel05a}.  This said, tests by \citet{narayanan11b} show
  that the thermal properties of the ISM in the molecular phase are
  relatively insensitive to these choices.

In order to simulate a diverse range of physical conditions,
   we concentrate in this work on major binary galaxy mergers, with
   total baryonic mass $M_{\rm bar} = 3.1 \times 10^{11}$.  The
   mergers are all identical on initialization, though vary in their
   orbits.  The physical properties of these galaxies are summarized
   in Table 1 of NK14, and in particular here we focus on models
   z0d4o, z0d4l, and z0d4e.  The varying orbital angle impacts the
   strength of the nuclear starburst upon final coalescence, and
   therefore the physical properties of the molecular ISM during the
   most heavily-star forming phases. 

The CO abundance depends on the carbon abundance, 
  set to $X_C = 1.5 \times 10^{-4}$ times solar metallicity, and the semi-analytic
  model of \citet{Wolfire2010} to determine the fraction of 
  carbon locked into CO. This fraction 
  varies by cell and simulation. Luminosity-weighted averages
  for each snapshot vary from about 25\% to 85\%.

\subsection{Determining Bulk Physical Conditions}

For each snapshot from the simulations, the physical properties of the
SPH particles were projected onto an adaptive mesh with an
  octree memory structure.  The neutral gas is assumed to all reside
  in giant, spherical, isothermal clouds of constant density.  The
  surface density is directly calculated from the mass within a given
  oct cell, though \citep[following ][]{narayanan11b} we consider a
  floor surface density of $\Sigma_{\rm cloud} = 85 M_\odot$
  pc$^{-2}$, comparable to observed values of local GMCs
  \citep{solomon87a,bolatto08a}.  The \htwo \ gas mass within these
  clouds is determined from the
  \citet{krumholz08a,krumholz09a,krumholz09b} formalism that balances
  the photodissociation rate of \htwo \ molecules by Lyman-Werner band
  photons against the growth rate of molecules on dust grains.  

\begin{figure}
\plotone{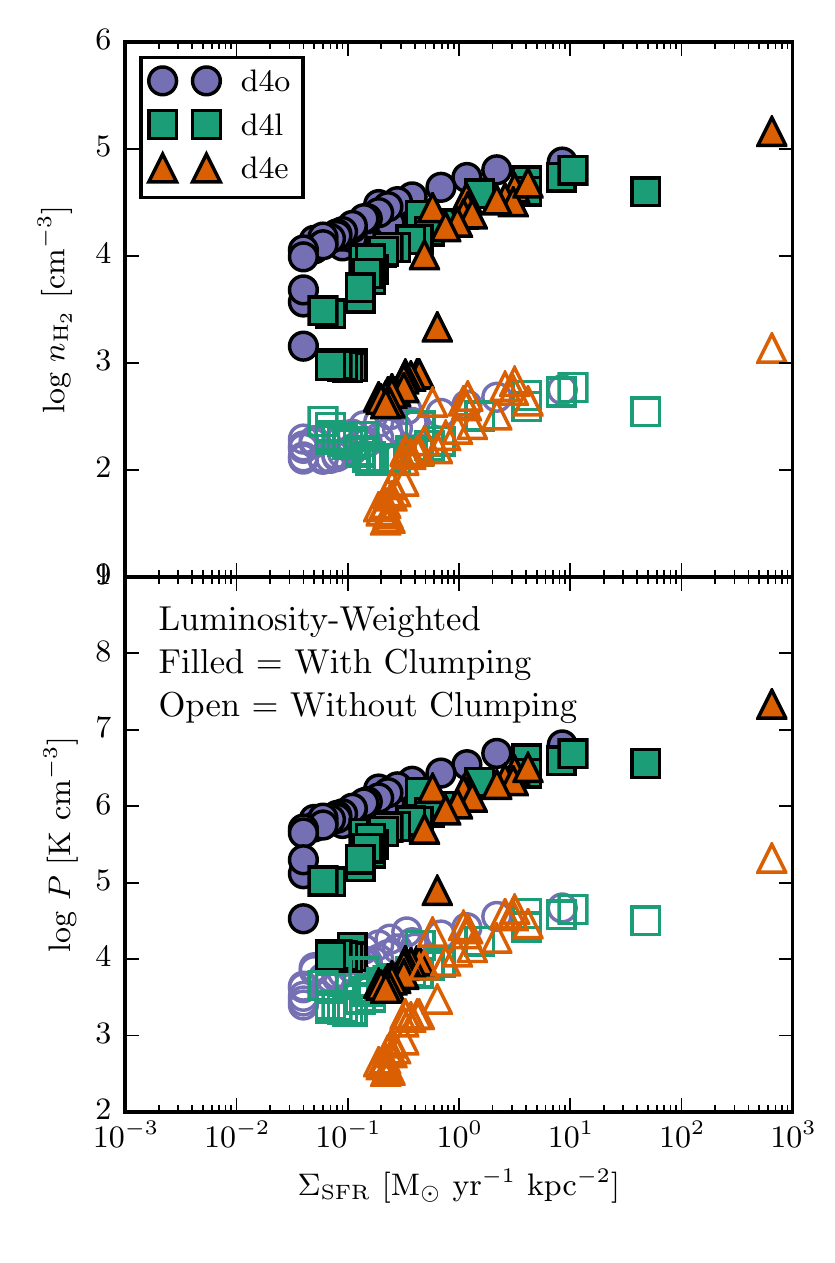}
\caption{Effect of clumping on gas density (and subsequently, pressure). When producing the SLEDs with DESPOTIC, the collision rate coefficients were enhanced (Equation \ref{eqn:clump}) due to clumping associated with the non-thermal velocity dispersion. The filled markers are the associated (higher) mass-weighted densities and pressures due to this clumping. We note that Figure 4 of NK14, which is similar in form to our Figures \ref{fig:masswvslumw} and \ref{fig:conditions_totlumlumweight}, do not show the effect of clumping. The non-clumped parameters in open markers are not used for subsequent analysis.}
\label{fig:clump}
\end{figure}

We model the sub-resolution turbulent compression (or ``clumping") of gas by
  scaling the volumetric densities by a factor $e^{\sigma_p^2/2}$,
  where $\sigma_p$ is a factor related to the 1 dimensional Mach
  number of the gas
  \begin{equation} \label{eqn:clump}
\sigma_p^2 \approx {\rm ln}\left(1+3M_{\rm 1D}^2/4\right) 
  \end{equation}
where this factor derives from turbulent box simulations
\citep{ostriker01a,padoan02a,lemaster08a}.  It is these re-scaled
densities that are used in calculating the collision rates for
excitation, and therefore these densities that we use for comparison
between simulations and mock observations. 
The effects on the densities are shown in Figure \ref{fig:clump}.
The total mass is conserved.

\begin{figure*}
\plotone{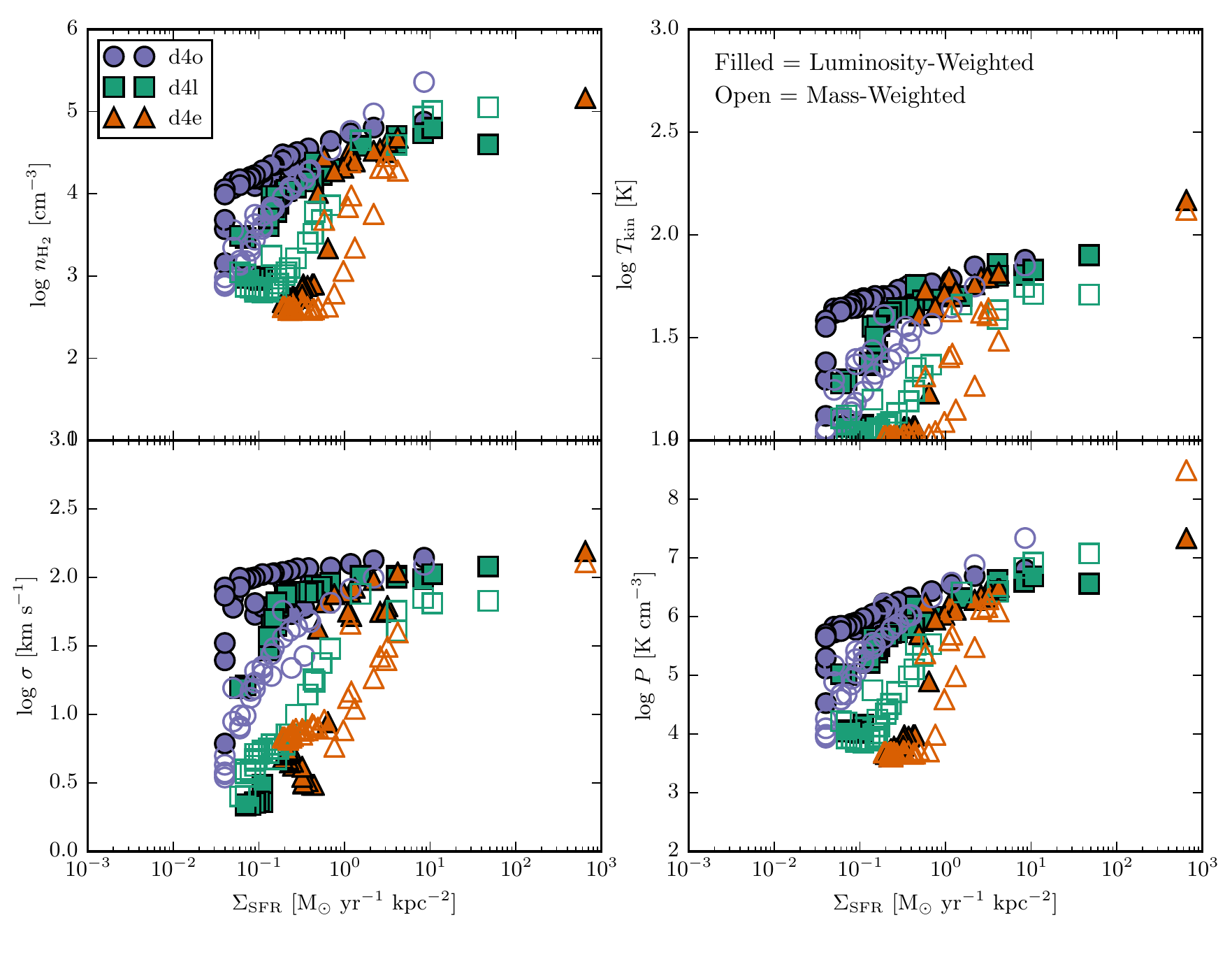}
\caption{Summary of the luminosity-weighted (filled) and mass-weighted (open) parameters from the SPH grids. The values of all parameters are generally higher when considering the luminosity-weighted versions; the difference is most enhanced for the low-$\Sigma_{\rm SFR}$ snapshots. 
For both types of weighting, the density (and therefore also the pressure) are enhanced due to a clumping factor derived from the velocity dispersion (see Figure \ref{fig:clump} and associated explanation).}
\label{fig:masswvslumw}
\end{figure*}

While the physical conditions are calculated for every cell,
  because, observationally, the bulk of SLED modeling is done for
  unresolved galaxies, we compare to weighted-averages of the physical
  properties (e.g. gas temperature, dust temperature, $\rho_{\rm
    H2}$) in our models.  We examine both the mass-weighted and CO-
  luminosity-weighted\footnote{Our CO-luminosity weighting is summed
    over the first 10 rotational transitions.} physical properties, and
  show these in Figure~\ref{fig:masswvslumw}. 

In each of these and subsequent figures, the different simulations are
indicated by different colors and marker shapes. Within a simulation,
different snapshots at different points in time correspond to
different star-formation surface densities. This quantity,
$\Sigma_{\rm SFR}$ is the x-axis our figures, to examine trends with
star formation activity. By visual comparison of the filled
(luminosity-weighted) and open (mass-weighted) markers, one can see
that the all of the parameters are generally higher when using the
luminosity-weighted values. The difference is most pronounced for the
low-$\Sigma_{\rm SFR}$ snapshots. 
The snapshots range from 5 to 10 Myr apart.

\subsection{Galaxy-Integrated SLEDs}

``Observed'' SLEDs were created for each simulation snapshot by taking
the integrated flux measurements of each CO transition with
$J_{upper}\leq10$ from \citet{Narayanan2014} and converting them into
a line brightness.  These calculations were done with {\sc despotic} \citep{Krumholz2013},
which operates under the escape probability formalism.  Here, the
thermal and radiative equilibrium are simultaneously solved for each model cloud.
For all CO transitions we assumed a total measurement error of $10\%$.

The galaxies were assumed to be unresolved and were placed at a fiducial redshift of $z=0.05$
and are intended to emulate SLEDs measured with Herschel or ALMA observations \citep[e.g.,][]{Kamenetzky2016,Lu2017}.
The SLED modeling requires an estimate of the size of the molecular gas; we used the area within a CO (1--0) contour of $1$ K km s$^{-1}$.
We also performed comparison fits using a luminosity weighted area (which is typically much smaller than the ``contour'' area), but the choice of area did not significantly alter our results for the physical conditions of the molecular gas.

\subsection{Line Fitting Procedure}
\label{sec:fittingproc}

The fitting of the ``observed" SLEDs follows the procedure in \citet{Kamenetzky2014}. We use the nested sampling algorithm \textsc{MultiNest} \citep{Feroz2009} and its python wrapper, \textsc{PyMultiNest} \citep{Buchner2014}, to compare the ``observed" SLEDs to those produced by the non-LTE code \textsc{RADEX} \citep{VanderTak2007}. As mentioned, the SLEDs themselves were produced using \textsc{DESPOTIC}, which introduces a minor inconsistency in our modeling.  We  address further in Section \ref{sec:results}. We use the same code as used to fit actual extragalactic SLEDs observed by the {\it Herschel}-SPIRE FTS reported in \citet{Kamenetzky2017}, \textsc{PyRadexNest} \citep{Kamenetzky2018}, available online\footnote{\url{https://github.com/jrka/pyradexnest}}. We also utilize the \textsc{Python} wrapper to \textsc{RADEX}, \textsc{PyRadex}\footnote{\url{https://github.com/keflavich/pyradex}}.

Each \textsc{RADEX} model depends on four free parameters: the kinetic temperature (\tkin), volume density of the collision partner with CO (molecular hydrogen, \nhtwo), column density of CO (\Nco) per unit linewidth, and the angular area filling factor ($\Phi < 1$), which linearly scales the fluxes produced by \textsc{RADEX}. The rotational level populations and optical depths of each line are iteratively determined, and then the intensities (as background-subtracted Rayleigh-Jeans equivalent radiation temperatures) are calculated using an escape probability method. We assume a background temperature of 2.73 K for the cosmic microwave background at $z=0$. In \textsc{PyRadexNest}, for a given set of parameters ${\bm p}$, we minimize the negative log likelihood of the predicted \textsc{RADEX} model $I({\bm p})$ given the measurements $\bm x$ and errors $\bm \sigma$ as:

\begin{equation}\label{eqn:likelihood}
-{\rm ln}(\mathscr{L})= \sum_i 0.5 {\rm ln}(2 \pi) + {\rm ln}(\sigma_i) + 0.5 (x_i-I_i(\bm{p}))^2 \sigma_i^{-2}. 
\end{equation}

In the above equation, $\mathscr{L}$ is the likelihood, $x_i$ is the ``measured" line intensity of a single CO transition, $\sigma_i$ is total uncertainty in a single transition measurement (10\% of $x_i$), and $I_i(\bf{p})$ the RADEX-modeled line intensity for that transition given the parameters ${\bf p} = [T_{\rm kin}, n_{\rm H_2}, N_{\rm CO},  \Phi]$, described in the preceding paragraph.

In practice, a few other galaxy-specific parameters are set in the modeling. We assume a linewidth of 
250 \kms; the total emission scales with this quantity, while the physical conditions 
(temperature, density) do not. 
The optical depth and escape probability in RADEX depend only 
on the column density {\it per unit linewidth}. Therefore, our total integrated 
line intensities must be divided by a linewidth for comparison to RADEX.
A different choice of linewidth 
therefore scales the total {\it integrated} emission, which one must use to calculate 
the {\it total} column density and then total mass. As we are only interested in temperature 
and density, our results do not depend on the choice of assumed linewidth for the line fitting procedure.

We also place three binary priors on the likelihood calculation. The prior is one if the simulated parameters satisfy all three conditions listed next to ensure physically plausible solutions, and zero if the simulated parameters violate any one condition. The first two conditions are an upper limit of $3\times10^{12}$ M$_{\odot}$ on the total mass and a maximum length of 10 kpc. Finally, due to the limits of the escape probability formalism used by \textsc{RADEX}, we only include lines with optical depths between -0.9 and 100 in the likelihood calculation. 

We focus primarily on what we will call the ``physical conditions" of the gas, namely the kinetic temperature and the density. Because the temperature and density are degenerate, we also focus on the pressure $P$/k = \tkin $\times$ \nhtwo. For each of these three parameters, we marginalize over all other parameters to find a one-dimensional probability distribution. From this distribution, we calculate a median value which represents our fitted estimate to compare to the grid value. We also calculate a $1\sigma$ width in the distribution to quantify the uncertainty in our fitted parameter.

\section{Results}\label{sec:results}

\begin{figure}
\includegraphics[width=0.45\textwidth]{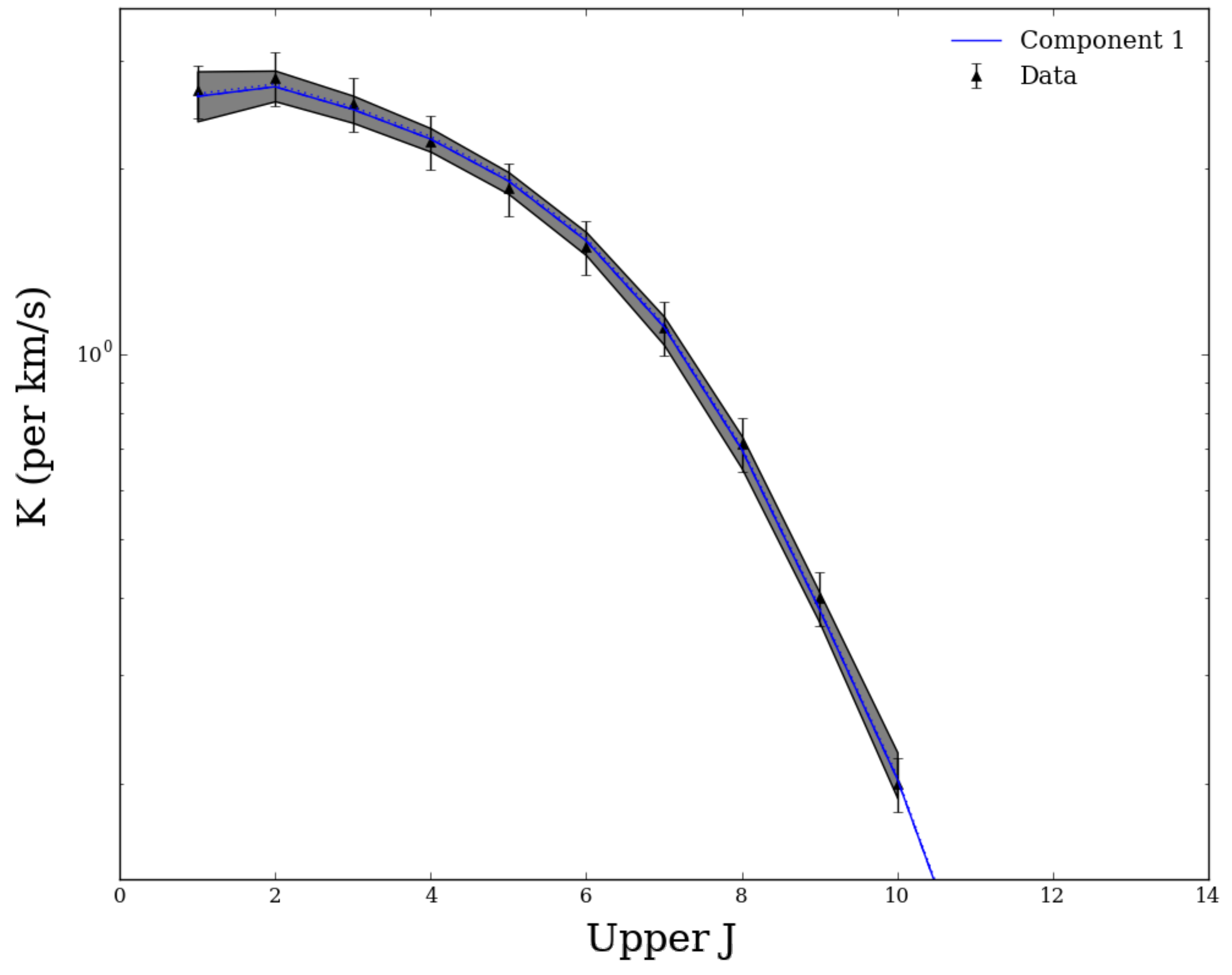}
\caption{
A sample CO SLED derived from the d4o simulation, showing the ``observed'' line fluxes and the best-fit model derived from the process described in Section~\ref{sec:fittingproc}. The straight blue line represents the best-fit SLED, and the shaded gray region represents the one-dimensional marginalized parameter distribution uncertainties for each individual line intensity modeled from RADEX.
CO SLEDs and best-fit models for the remainder of the simulations and snapshots are shown in Figures~\ref{fig:sleds_d4e} and \ref{fig:sleds_d4o}.}
\label{fig:sampleSLED}
\end{figure}

In Figure~\ref{fig:sampleSLED} we show a representative CO SLED, derived from the d4o simulation, along with the best-fit single-component model and uncertainties obtained following the procedure outlined above.
The ``observed'' SLED is well-fit by the model, through the \jten\ line.
Below we describe the differences between the simulated SLEDs and real galaxy SLEDs and how the best-fit model parameters correspond to the ``true'' values of the hydrodynamic simulations.

\subsection{Differences Between Simulated and Real SLEDs}

Our model SLEDs exhibit a number of features that are unlike those observed in real galaxies. 
We find that our model SLEDs (through \jten) are well-fit by one component of gas, whereas real galaxy-integrated SLEDs, 
such as those in \citet{Kamenetzky2017}, require two components. 
In real galaxies, when combining ground-based data (\jone\ through usually \jthree) and SPIRE FTS line measurements (\jfour\ through \jthirteen), the SLED is not well described by a single component of gas; the emission of the low-J lines is largely from cold gas, which falls off quickly by mid-J (\jfour\ through \jsix) lines. 
A second, warmer component of gas is responsible for the emission of the mid-J lines and higher. 
When we tried to fit our model SLEDs with two components of gas, we found that the statistically best-fit SLED was one component anyway (with the second component being unconstrained, so long as it contributed negligibly to the fit).

Similarly,  unlike real galaxy-integrated SLEDs, we find small uncertainties in the marginalized parameters (temperature, density, pressure) when modeling as a single component. Although we include 10\% error on the ``observed" data points, the smooth behavior of the simulated SLED is often uniquely fit by a small set of parameter combinations. 
The median uncertainty for the density, temperature, and pressure was 0.2, 0.03 and 0.2 dex, respectively. For comparison, \citet{Tunnard2016} found the recoverability of single component \textsc{RADEX}-created SLEDs to be about 0.5 dex without using isotopologues. In the two-component models of \citet{Kamenetzky2017}, the warm component of gas is the best constrained and most comparable to this work; the uncertainty in the warm component pressure was about 0.3 dex (but 1.0 dex for the cold component; when modeled as two components, there is a larger degeneracy between parameters).

\subsubsection{Comparison to Luminosity Weighted Parameters}

\begin{figure*}
\plotone{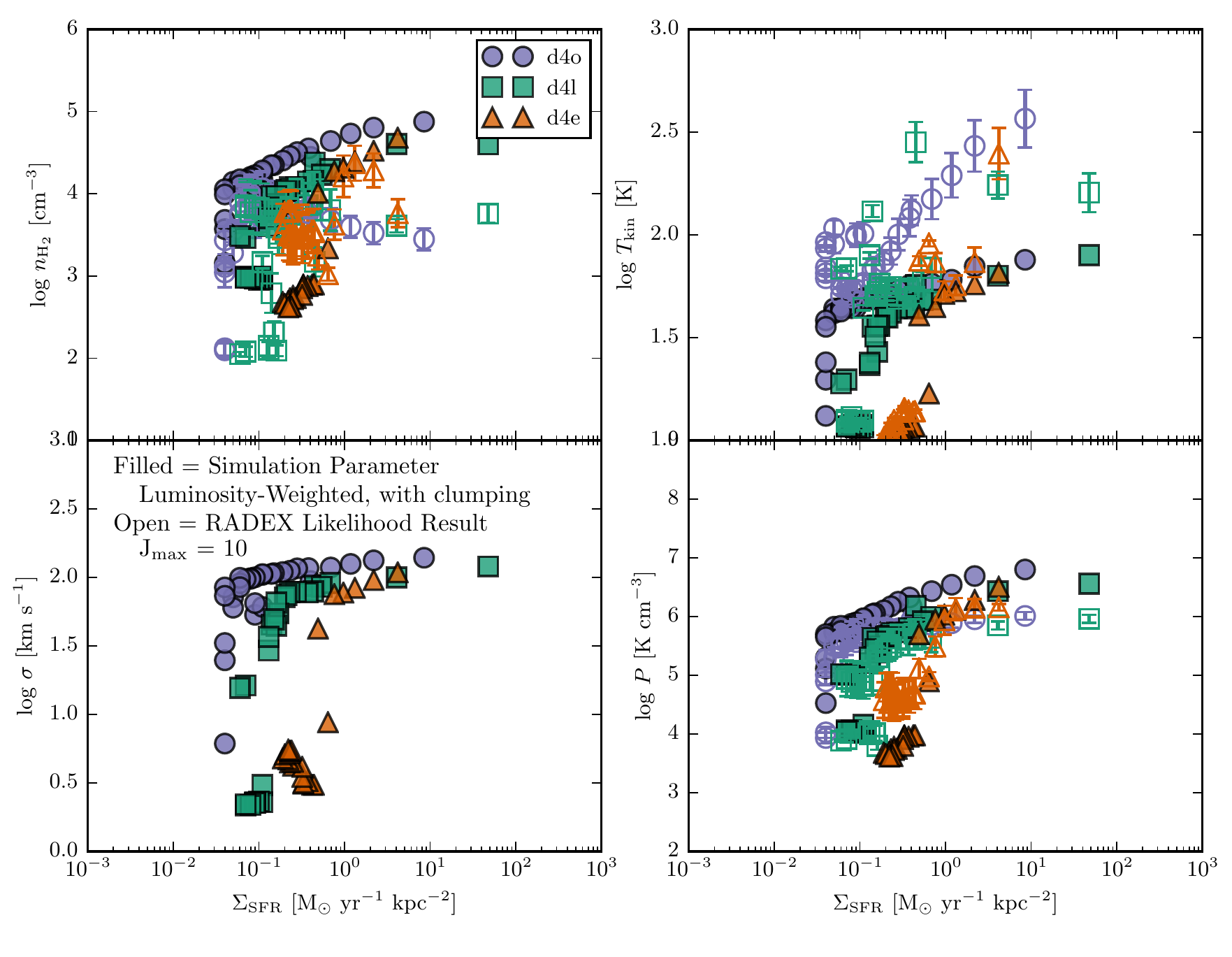}
\caption{Comparison of the luminosity-weighted vs. likelihood fitted parameters. Filled symbols are the simulation parameters; each simulation has a specific color and symbol. Open symbols of equivalent color/shape correspond to the \textsc{PyRadexNest} likelihood results. }
\label{fig:conditions_totlumlumweight}
\end{figure*}

Figure \ref{fig:conditions_totlumlumweight} compares the luminosity-weighted simulation parameters (filled symbols) to the \textsc{PyRadexNest} derived parameters from fitting the ``observed" SLEDs (open symbols). As was shown in Figure \ref{fig:masswvslumw}, had we used the mass-weighted simulation parameters, the parameters (especially density) would be even lower and show a greater discrepancy between the ``observed" parameters. Therefore, we focus on the luminosity-weighted parameters. However, there are still notable discrepancies, which we now investigate further.

\begin{figure*}
\plotone{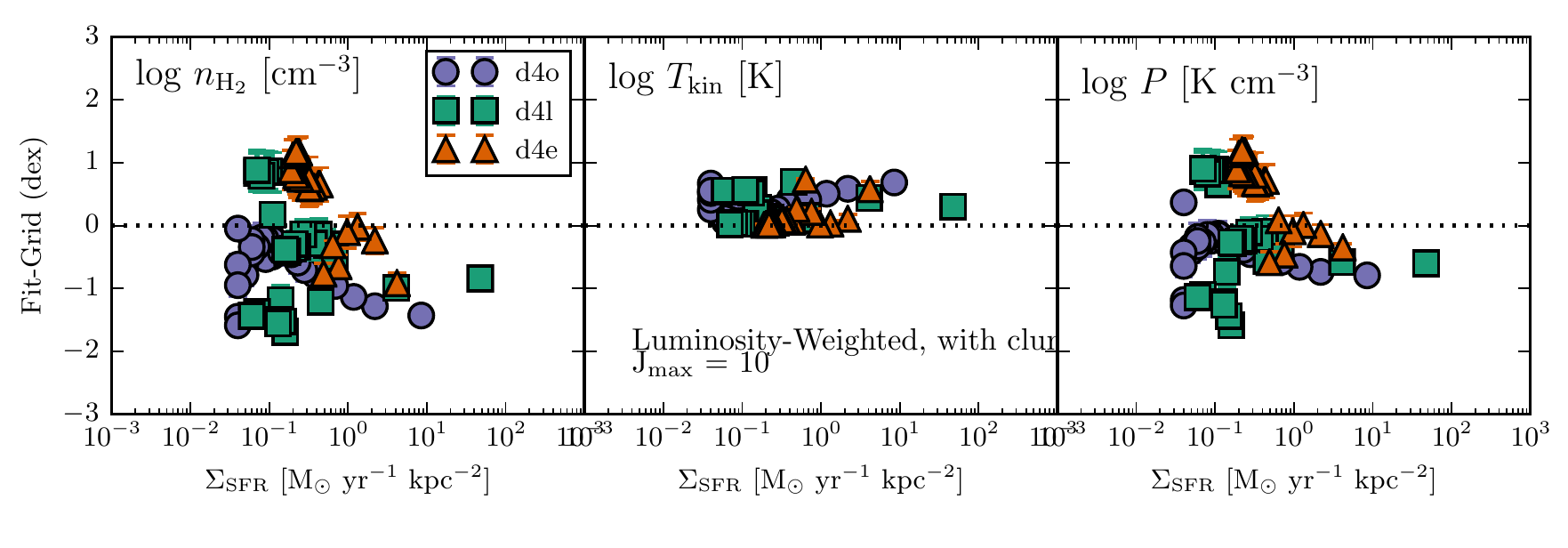}
\caption{Differences between the luminosity-weighted vs. likelihood fitted parameters. }
\label{fig:conditions_totlumlumweight_diff}
\end{figure*}

Figure \ref{fig:conditions_totlumlumweight_diff} shows the difference in fitted vs. luminosity-weighted parameters for density, temperature, and pressure. The three simulations (shown in different colors), which have different merger properties, do not follow the same trends. Each one appears to have trends with $\Sigma_{\rm SFR}$ as the snapshots evolve over time. On this difference plot of \textsc{PyRadexNest} fitted value minus CO-luminosity weighted simulation parameters, data points above the dashed zero line indicate snapshots for which our fitted values are higher than the grid values. For the most part, our fits underestimate the density, match or slightly overestimate the kinetic temperature, and underestimate the pressure. In some snapshots, however, we overestimate density and pressure instead. 

\begin{figure*}
\plotone{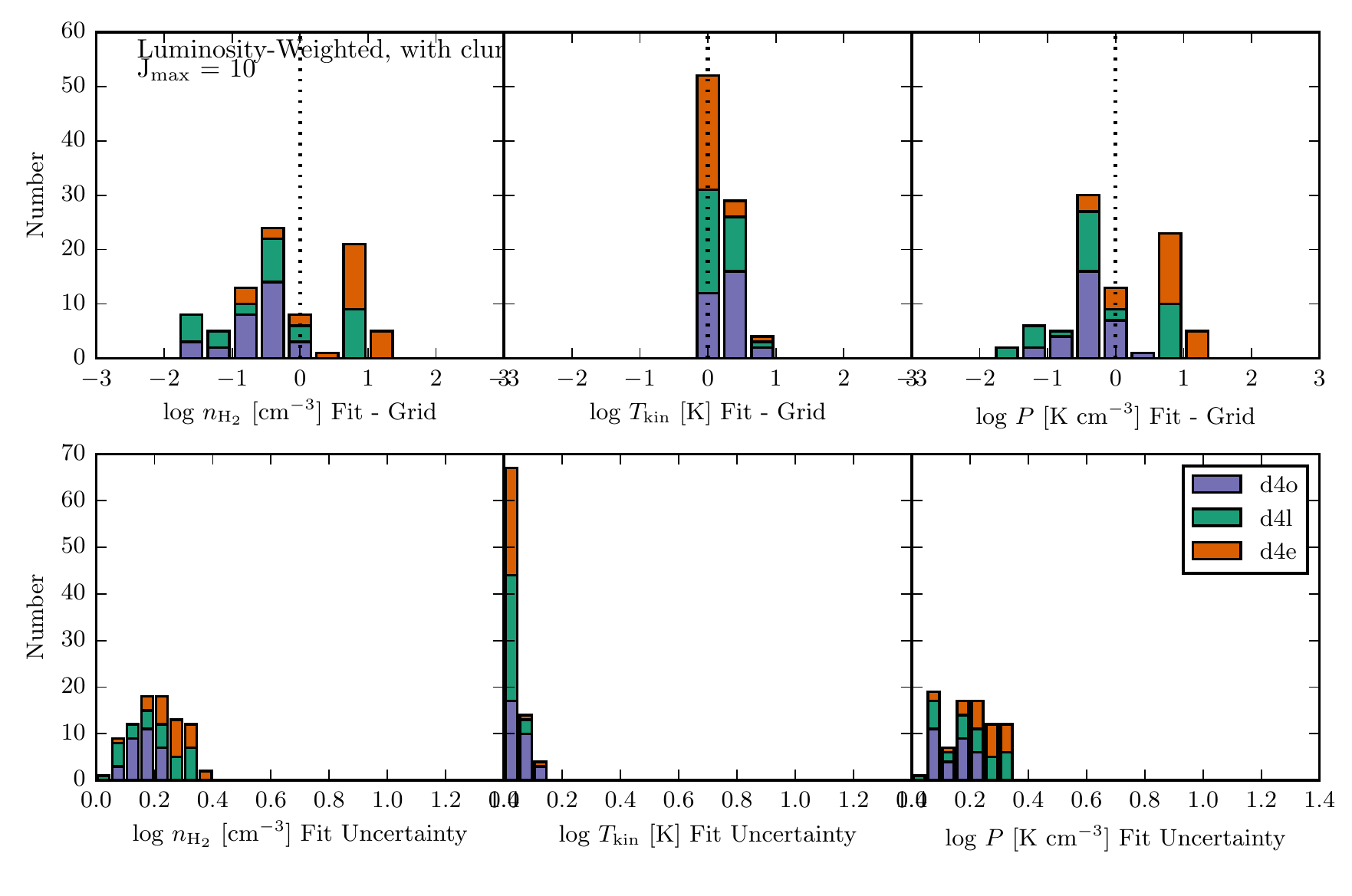}
\caption{Top row: Histogram of the difference between the luminosity-weighted and likelihood fitted parameters.
Bottom row: Uncertainty (in dex) in the fitted parameters. }
\label{fig:conditions_totlumlumweight_hist}
\end{figure*}

In Figure \ref{fig:conditions_totlumlumweight_hist}, we show
histograms of the differences between the luminosity-weighted and
likelihood fitted parameters. The fitted, statistical uncertainties
themselves do not take into account the differences between using
\textsc{Radex} for fitting and \textsc{Despotic} for the creation of
the SLEDs.

The \textsc{PyRadexNest} best fit models are highly weighted by and sensitive to the high-J line luminosity. We also examined the impact of comparing to simulation parameters weighted by each grid point's \jeight\ luminosity instead of total CO-luminosity. Overall, the distributions are similar to those shown in Figures \ref{fig:conditions_totlumlumweight}, \ref{fig:conditions_totlumlumweight_diff}, and \ref{fig:conditions_totlumlumweight_hist}. However, for some of the lowest $\Sigma_{\rm SFR}$ snapshots ($< 0.1$ \ms yr$^{-1}$ kpc$^{-2}$), the CO \jeight\-luminosity weighted parameters are slightly higher than the total CO-luminosity weighted parameters, and are better matched by our \textsc{Radex} models. 

\subsubsection{Dependence on Number of Lines Modeled and Area}

\begin{figure*}
\plotone{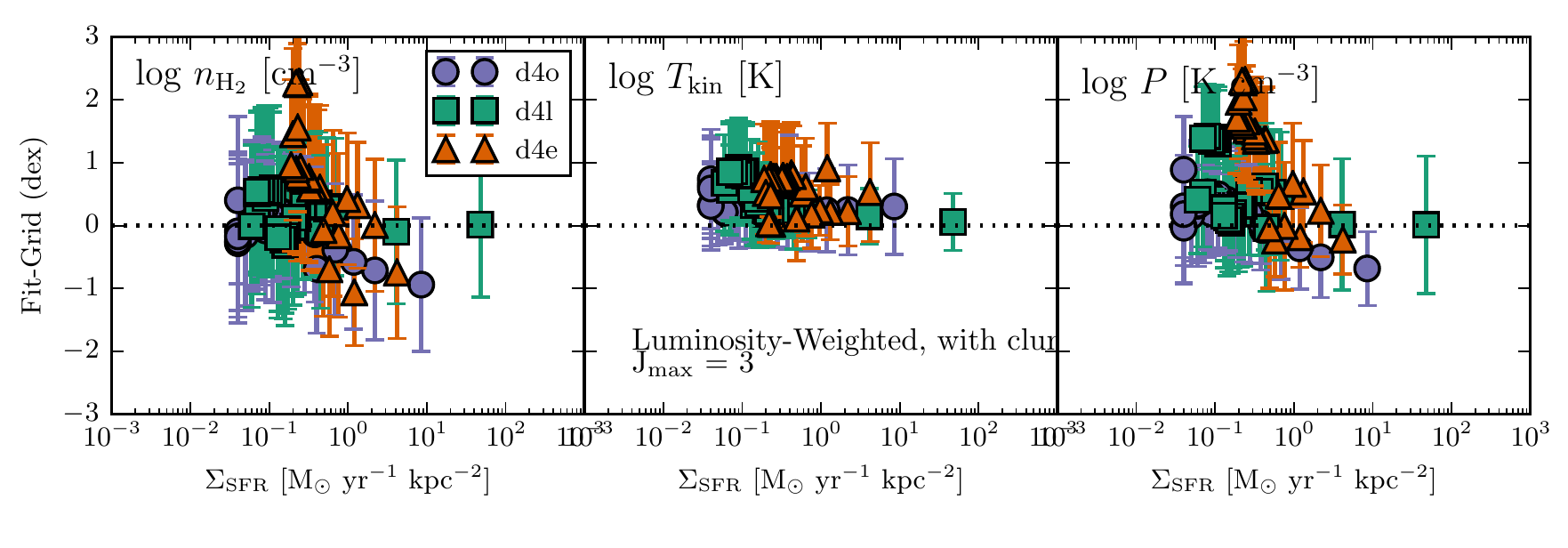}
\caption{Differences between the luminosity-weighted vs. likelihood fitted parameters, for only modeling up to \jthree.}
\label{fig:conditions_lumweight_diff_maxJ3}
\end{figure*}

\begin{figure*}
\plotone{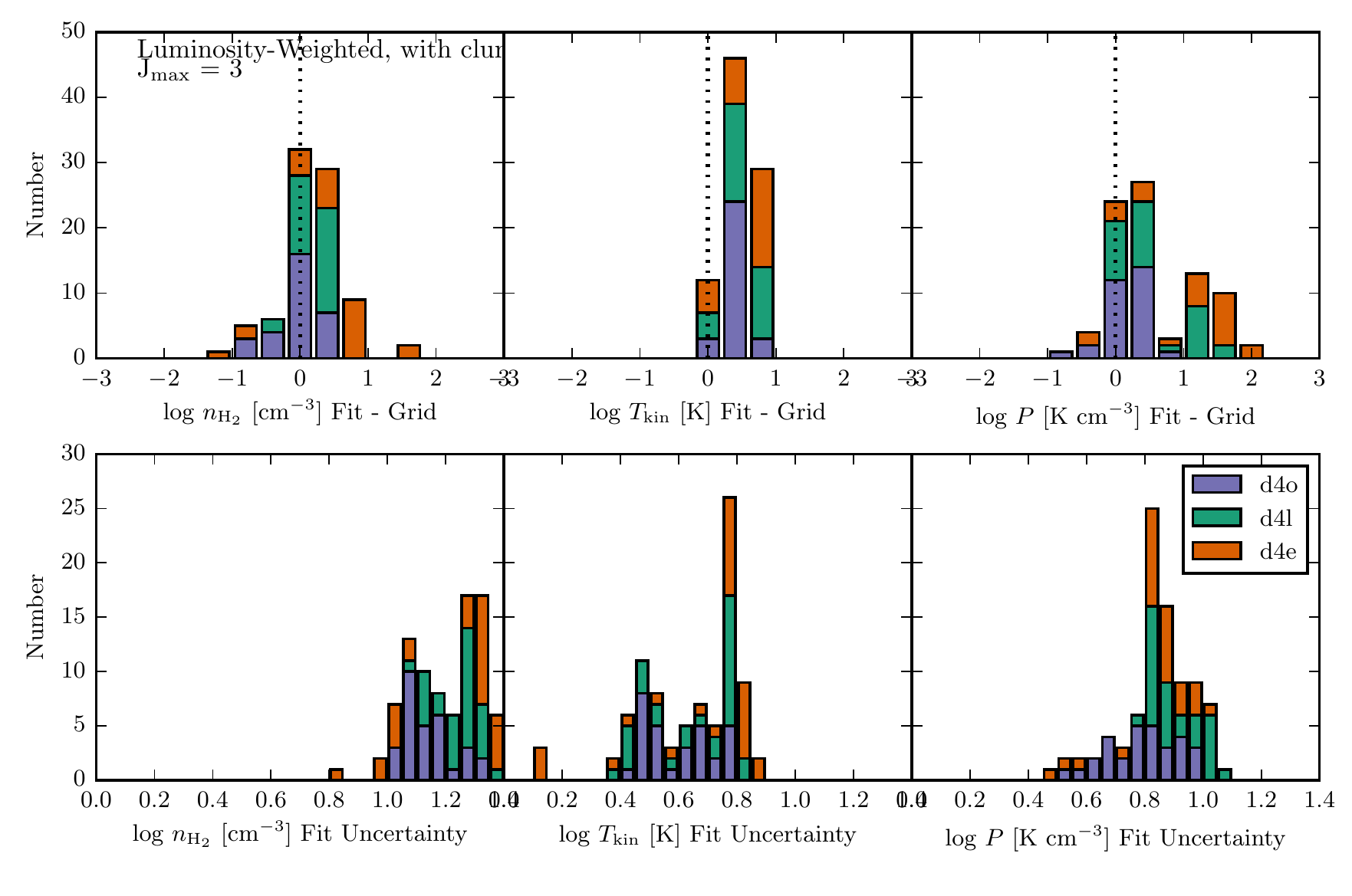}
\caption{Top row: Histogram of the difference between the luminosity-weighted and likelihood fitted parameters, for only modeling up to \jthree.
Bottom row: Uncertainty (in dex) in the fitted parameters.}
\label{fig:conditions_lumweight_hist_maxJ3}
\end{figure*}

We also compared the values derived from modeling only up to \jthree\ instead of \jten. The result is much higher uncertainties in the parameters, but median values which align better with the luminosity-weighted parameters (see Figures \ref{fig:conditions_lumweight_diff_maxJ3} and \ref{fig:conditions_lumweight_hist_maxJ3}). This is likely because most of the CO luminosity, even in simulation grid cells with extreme conditions, is in the low-J lines. 

In our fitting algorithm, like in many others, the likelihood of a model \textsc{Radex} SLED is weighted by the absolute value of the error bar of each data point. Assuming a constant relative error on each data point (e.g. 10\% here) means that the relatively low-luminosity, high-J lines are significantly more heavily weighted in the fit. (Figures \ref{fig:sleds_d4e} and \ref{fig:sleds_d4o} demonstrate the large dynamic range of the SLEDs.) Fitting the whole SLED as one component seems to drive the fitted parameters to more diffuse, slightly hotter gas. Fitting only the low-J lines drives the fitted parameters to more dense, cooler gas that are more representative of the total luminosity-weighted parameters. 

\subsubsection{Dependence on Area}

The mass depends on the product of the total column density and the assumed area of emission. If we change the area, the mass scales accordingly, but it does not significantly affect the physical conditions (temperature, density) because they are much more dependent on the shape of the SLED.

\subsection{RADEX vs. DESPOTIC}

\begin{figure*}
\plotone{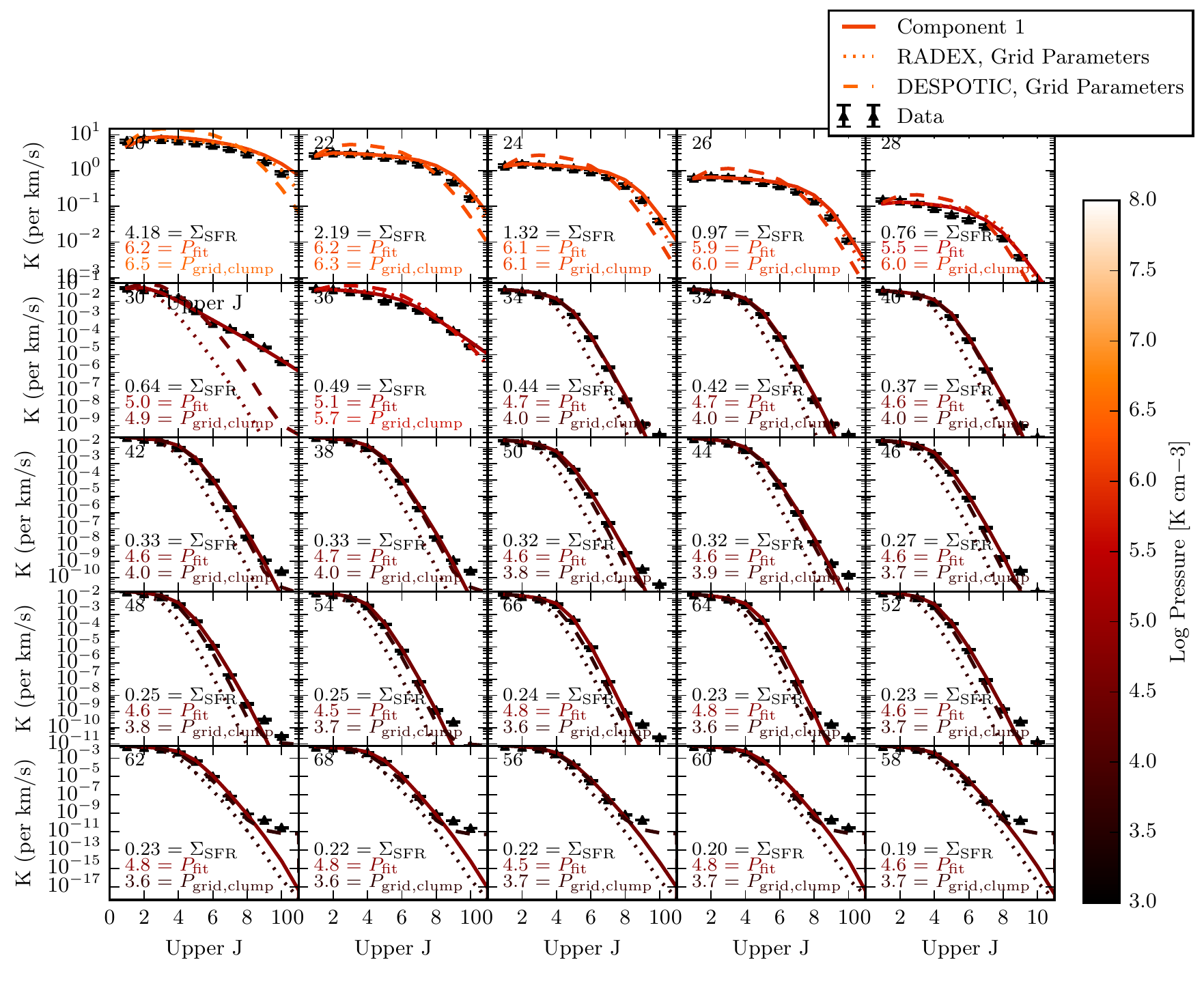}  
\caption{Spectral Line Energy Distributions (SLEDs) for snapshots of d4e. Snapshots are in order from highest (upper left) to lowest (lower right) $\Sigma_{\rm SFR}$.
Each panel shows the galaxy-integrated SLED and the best-fit solution (solid line, color coded by pressure). 
We also show the RADEX and DESPOTIC (dotted, dashed) SLEDs using the luminosity-weighted temperature and density (effective clump density for RADEX) instead.
We use the same column density and filling factor from the likelihood results.
We fix them to the same \jone\ value as the best fit to better see the relative shapes.
}
\label{fig:sleds_d4e}
\end{figure*}

\begin{figure*}
\plotone{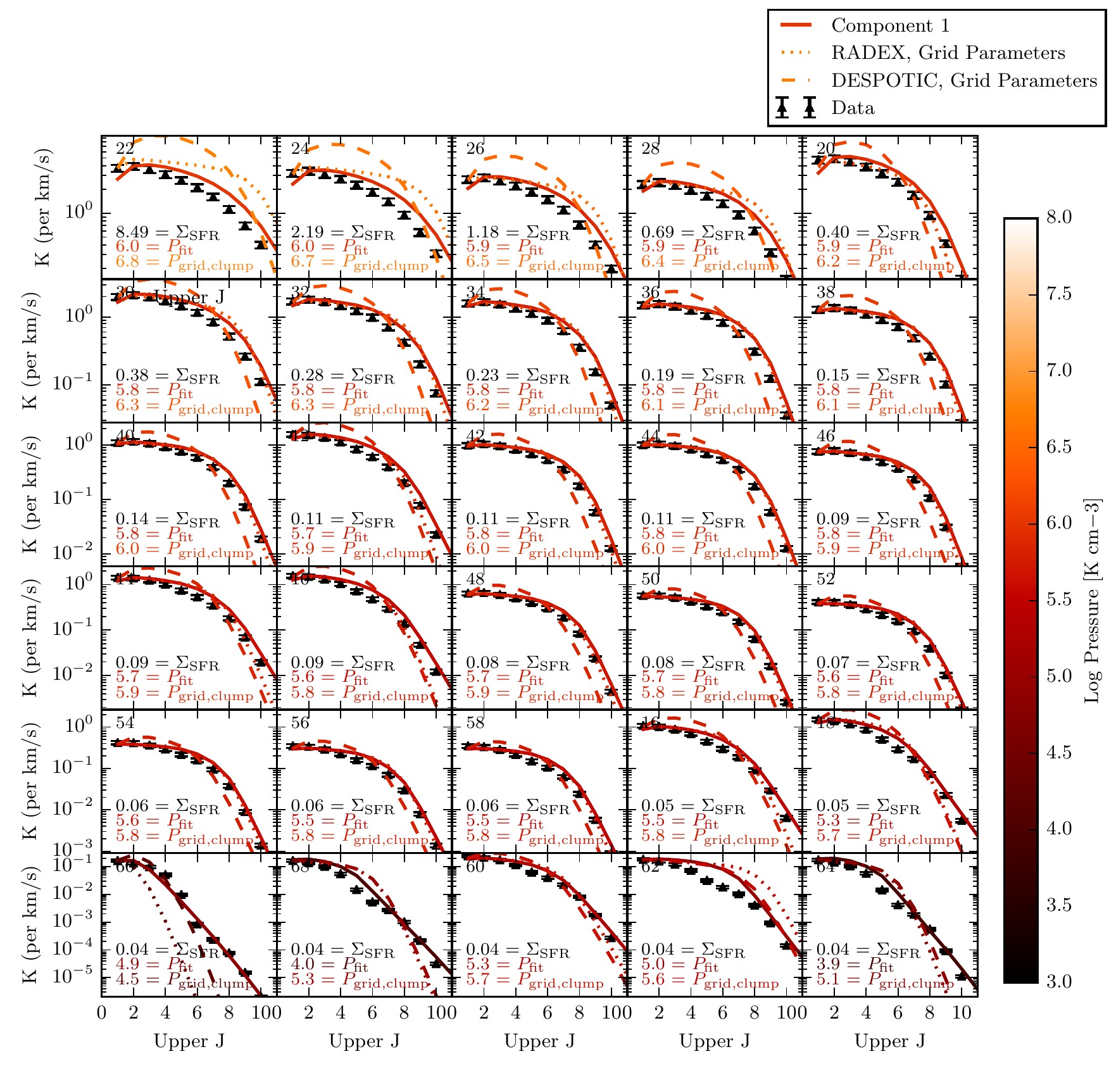}
\caption{Spectral Line Energy Distributions (SLEDs) for snapshots of d4o. See caption of Figure \ref{fig:sleds_d4e}.}
\label{fig:sleds_d4o}
\end{figure*}

In Figures \ref{fig:sleds_d4e} and \ref{fig:sleds_d4o}, we show the galaxy-integrated ``observed SLEDs", the best-fit results from RADEX likelihood fitting, and the RADEX and DESPOTIC SLEDs that correspond to the temperature and density from the grid-weighted parameters.

For most of the lower $\Sigma_{\rm SFR}$ snapshots of d4e (Figure \ref{fig:sleds_d4e}), the main differences in the grid and fit results seem to be difference between RADEX and DESPOTIC (because the dashed DESPOTIC lines from the grids match the best-fit from RADEX). For d4o (Figure \ref{fig:sleds_d4o}), however, we find the galaxy-integrated SLEDs are not particularly consistent with either the RADEX or DESPOTIC SLED corresponding to the luminosity-weighted parameters.  Appendix D of \citet{Krumholz2013} provides a detailed comparison to RADEX.

\section{Discussion}\label{sec:discussion}

Fundamentally, we and others are attempting to answer the question: 
how good are our modeling techniques at determining the bulk physical 
properties of a complex ensemble of molecular gas? Even the highest 
resolution maps of nearby galaxies must convolve together a vast range 
of gas densities, temperatures, and dynamical properties. For all but the 
closest galaxies, line emission ratios must be constructed using a single 
integrated beam measurement or a map of only a handful of beams 
spanning the entire galaxy. We have chosen to model the lowest-resolution 
(one beam) scenario.

There are really two separate aspects (precision and accuracy) to the 
aforementioned question: 1) what are the inherent uncertainties 
in the modeling techniques, and 2) how accurately do the median quantities 
represent the actual quantities? To the first question, the work by \citet{Tunnard2016} 
nicely showed that one should consider median quantities from LVG models 
(kinetic temperature, volume density, and velocity gradient) uncertain to at least 0.5 dex.
Their simulated observational data included an optimistic uncertainty of 10\%, the same as 
we do here, though they use three more lines than we do here (up to \jthirteen).
Our approach differs from \citet{Tunnard2016} 
in that we model the SLED integrated over all the cells in the simulation; 
each individual grid cell of the hydrodynamic simulation has its own smooth SLED.
These integrated SLEDs, which are also smoothly varying, can be modeled by one component.
Our approach also differs because  \citet{Tunnard2016}
introduced slight errors on the lines produced by RADEX to better emulate real observations. 
We modeled the smoothly-varying DESPOTIC-produced SLEDs without any added noise. 
This likely caused our integrated parameter uncertainties to be slightly smaller (bottom panel of Figure 
\ref{fig:conditions_totlumlumweight_hist}). 
Our uncertainties of approximately 0.2 dex should be considered the 
``best-case scenario" when ten or more lines are available.

The precision drops significantly when only low-J lines are used (bottom panel of 
Figure \ref{fig:conditions_lumweight_hist_maxJ3}). 
A vast range of physical conditions can produce the same low-J line emission.
The distinguishing feature of the SLED is the point at which it turns over, usually in the mid-J 
region of our range. Without determining the approximate point of turnover and 
slope of the SLED after turnover (when plotted in luminosity units vs. $J$-line), 
one cannot precisely determine the 
bulk physical properties of the gas. Prior to ALMA and {\it Herschel}, this was the 
case for the majority of galaxies. Some high-J lines 
were available for bright galaxies under the best conditions \citep[e.g. CO \jsix\ 
from JCMT, ]{Papadopoulos2010}.
The approximately uncertainties in the density, 
temperature, and their product (pressure) of 1.0-1.4, 0.4 - 0.8, and  0.8 - 1.0 dex, respectively, 
should be considered the ``worst-case scenario" when only low-J lines are available.

Nearby galaxies observed with {\it Herschel} contain enough lines for modeling up to \jthirteen, but 
often require at least two unique components to fit the SLED, as done in \citet{Kamenetzky2017}. 
With two components comes a total of eight free parameters in the models; there is degeneracy 
between the cold and warm components parameters. The resulting uncertainties in the models, 
however, largely resemble what we find here \citep[Figure 1 of ][]{Kamenetzky2017}. The warm 
component, which is largely fit by the high-J lines, has a pressure uncertainty of about 0.3 dex. 
The cool component, which is largely fit by the low-J lines and is analogous to our low-J only models here, 
are uncertain to about 1.0 dex. The molecular mass, using a variety of methods, was found to be 
uncertain to a factor of about 0.4 dex on average (we do not focus on the molecular mass in this work). 
The uncertainty is larger, and result systematically offset (low), if the CO \jone\ line is absent, 
because the majority of the molecular mass is present in the ground state. 

The recent progress made in the area of submillimeter observations of high-redshift galaxies offers 
a new, different challenge. A full CO SLED from \jone\ to \jten\ or \jthirteen\ is rare and often difficult 
\citep{Carilli2013,Casey2014a}. 
For galaxies in the redshift range $\sim0.3-1.5$ the \jone\ line is not accessible to sensitive facilities such as ALMA or the VLA, making the estimation 
of mass particularly uncertain. Which lines are available from ground-based observatories such 
as ALMA depend sensitively on the redshift. If at least a few lines are available that somewhat span 
the range from \jone\ through \jthirteen\ (for example, a SLED with \jtwo, \jfive, \jseven, which will likely
encompass the SLED's turnover), 
the uncertainties in the physical conditions would likely be bracketed by our best-case scenarios (0.2 dex) 
and worst-case scenarios (1.0 dex). Of course, the best way to determine the parameters' uncertainties 
is to examine the relatively likelihoods over a large parameter space using a nested-likelihood algorithm 
like we do here with MultiNest, or a Markov chain Monte Carlo (MCMC) method.

The second aspect of our question, that of accuracy, is harder to answer. What does it mean 
to accurately determine the bulk properties of a complex range of molecular gas clouds 
spanning an entire galaxy? As summarized in \citet{Leroy2017}, the ``density" of gas could refer 
to the collider density, critical density, effectively critical density (taking into account radiative 
line trapping), most effective density for emission, median density for emission, or median density 
by mass. Their modeled emission takes into account a realistic sub-resolution distribution of densities 
using log-normal and power law distributions, but fix the temperatures and optical depths of all 
gas clouds to the same value. Even for a fixed temperature and optical depth, the emission of a molecular 
line varies with density. As \citet{Leroy2017} points out, regions with lower densities can still emit, but with 
lower efficiency. Lower-density gas can contribute significantly to the total galaxy-integrated emission if it 
is present in a large enough amount. Our galaxy evolution simulations are more detailed in that they 
allow all properties (mass, temperature, density, velocity dispersion) to vary on a cloud-by-cloud basis. 

A clear conclusion from our work is that modeling of galaxy-integrated SLEDs does not accurately reproduce 
mass-weighted quantities, which are significantly lower in density, temperature, and velocity dispersion 
(Figure \ref{fig:masswvslumw}). Bulk properties derived from SLED fitting more accurately describe 
luminosity-weighted quantities. We find systematic offsets by property. For the gas density, 
our fitted parameters are systematically low (but not always, Figure \ref{fig:conditions_totlumlumweight_diff}).
For the kinetic temperature, our fits are slightly systematically high. Both of these parameters are degenerate 
with one another, but their product (pressure) is often better determined. Our resultant pressures 
are much closer to the mass-weighted pressures from the simulation and follow the same general trend 
of increasing with $\Sigma_{\rm SFR}$ (Figure \ref{fig:conditions_totlumlumweight}, bottom right). When using
only the low-J lines, our median properties are systematically closer to the luminosity-weighted properties of the 
simulations, but as discussed previously, the uncertainties were much higher.

This demonstrates that the beam-integrated emission from galaxies is dominated by the brightest, most 
extremely excited molecular gas. Such highly excited gas represents a small fraction of the total mass, consistent with 
the findings of \citet{Kamenetzky2014}.
For only the smallest $\Sigma_{\rm SFR}$ snapshots studied here was there a difference between properties weighted by \jeight\ vs. 
total CO luminosity. For these scenarios, the high-$J$ emission alone greatly weights the total integrated SLEDs.


\section{Conclusions}\label{sec:conclusions}

Any CO SLED integrated over a large area is the sum of a gradient in physical conditions (temperature, density). 
Given a large number of free parameters for each component of gas (temperature, density, column density, area filling factor) and often a small number of molecular line luminosities available for fitting, observers must necessarily model the smallest number of components to make statistically robust conclusions. 
These components (usually one, two, or occasionally three) are an oversimplification of a complex galactic system.
We sought to take a computational model of such a complex galactic system and ``compress" its information into one total integrated SLED, as an observer would see, and then model the gas as observers do. Our main conclusions are as follows:

\begin{enumerate}
\item When fitting CO SLEDs as a discrete number of components, the resultant parameters should be considered analogous to luminosity-weighted parameters, not mass-weighted. The highest luminosity regions of galaxy SLEDs represent the most excited conditions, but a small fraction of the mass.

\item For large $\Sigma_{\rm SFR}$ snapshots, the luminosity-weighted parameters (temperature, density, and pressure) were the same whether we weighted by CO \jeight\ or the total CO-luminosity.  For small $\Sigma_{\rm SFR}$, however, weighting by CO \jeight\ resulted in slightly higher temperatures, densities, and pressures, indicating that high-J emission has a greater influence on the SLED when $\Sigma_{\rm SFR}$ is low

\item When only using low-J lines (\jone, \jtwo, and \jthree), the uncertainties in the derived physical quantities are approximately one order of magnitude. The true luminosity-related quantities generally fall within the range of uncertainty. 

  \item On the other hand, when fitting the first $10$
    rotational lines, the uncertainty is usually about $0.2$ dex, through the true
    luminosity-weighted densities, temperatures and pressures generally fall outside this range of uncertainty. 
    This indicates a systematic difference between our recovered properties and the true luminosity-weighted properties, though they are close. 
     An uncertainty of $0.2$ dex is likely to be a lower limit on the uncertainty as SLEDs are rarely more well sampled than with data covering the first $10$ rotational lines.

    \item We therefore suggest that the typical systematic
      uncertainty on the physical properties when SLED modeling lies
      between $0.2-1$ dex., depending on the number of lines modeled, the sampling of the SLED in energy space, and the uncertainties of the integrated line fluxes.

\end{enumerate}

\acknowledgments

J. K. was supported by the National Science Foundation under Grant
Number AST-1402193. 
G.C.P. acknowledges support from a FONDECYT Postdoctoral Fellowship (No. 3150361)? and from the University of Florida.
 G.C.P. thanks the Sexten Center for
Astrophysics (http://www.sexten-cfa.eu) where part of this work was
performed. Partial support for D.N. was provided by NASA HST
AR-13906.001, HST AR-15043.0001, NSF AST-1724864 and NSF AST-1715206.
The Cosmic Dawn Center is funded by the Danish National Research
Foundation.

\software{RADEX \citep{VanderTak2007}, PyRadex (\url{https://github.com/keflavich/pyradex}), MultiNest \citep{Feroz2009}, PyMultiNest \citep{Buchner et al. 2014}, PyRadexNest \citep{Kamenetzky2018}, DESPOTIC \citep{Krumholz2013}}

\bibliography{./dn.bib,./jk.bib,./gcp.bib}
\end{document}